\documentclass[12pt,twoside]{article}
\usepackage{amssymb}
\usepackage{amsmath}
\usepackage[makeroom]{cancel}
\usepackage{latexsym}
\usepackage{longtable}
\usepackage{epsfig}
\usepackage{graphicx,bbm,psfrag}
\graphicspath{{images/}}

\makeatletter
\def\blfootnote{\xdef\@thefnmark{}\@footnotetext}
\makeatother

\def\LL{{\cal L}}

\def\CC{{\cal C}}

\def\0{\emptyset}

\def\a{\alpha}
\def\n{\nu}
\def\b{\beta}
\def\m{\mu}

\def\e{\varepsilon}

\def\o{\omega}

\newcommand{\pa}{\partial}

\def\be{\begin{equation}}
\def\ee{\end{equation}}
\def\bea{\begin{eqnarray}}
\def\eea{\end{eqnarray}}

\def\nn{\nonumber}

\begin{document}
\noindent
\textit{From Newton to Boltzmann: Hard Spheres and Short-range Potentials}, by Isabelle Gallagher, Laure Saint-Raymond and Benjamin Texier, Z\"{u}rich Lectures in Advanced Mathematics, vol. 18, European Mathematical Society, Z\"{u}rich, 2014, xi+135 pp., ISBN 978-3-03719-129-3.\blfootnote{\vspace{1mm}}\\\\\\

During the second half of the 19th century it became a widely accepted scientific view that matter consists of atoms/molecules and that their motion is governed by the same dynamical laws as the trajectories of heavenly bodies.
Of course, there was no knowledge on the precise form of the interaction potential. One relied on a model
of very tiny balls which interact through elastic collisions (hard spheres), perhaps  improving phenomenologically  
to a smooth interaction potential of short range. The necessity of the more realistic Lennard-Jones type potential was recognized only considerably later on the basis of, at the time, very precise thermodynamic measurements.

In 1872 Ludwig Boltzmann started from the then accepted atomistic model and wrote down a nonlinear transport equation for the one-particle distribution function. In modern terms, he argued that at a collision the two incoming velocities are statistically independent, which leads to a closed equation for the Boltzmann $f$ function. Such assumption could hold approximately only for a rarified gas. In the  6th problem of his famous 1900 Paris address,  D. Hilbert explicitly mentions Boltzmann's limiting processes. But mathematical progress was slow. Around 1950 H. Grad \cite{G49} first proposed 
a definite limit in which the Boltzmann equation would become exact, hence Boltzmann-Grad limit.
One considers a fixed box $\Lambda \subset \mathbb{R}^3$ containing $N$ hard spheres of radius $\e$
and has to study the limit where $N \to \infty$, $\e \to 0$ such that the mean free path $|\Lambda|/ \e^2N$
remains constant. In this limit the volume filled by hard spheres tends to 0 proportional to $\e$,
hence physically the fluid is at low density.
C. Cercignani \cite{C72} noted that Grad's limit fits very nicely with the structure of the BBGKY hierarchy for hard sphere correlation functions\footnote{The original version of the hierarchy was written for smooth potentials and
appears first in the papers by Bogolyubov, Born, Green, Kirkwood and Yvon, see e.g. \cite{Bog62}.}. 
With this insight, at the 1974 Battelle Rencontres, Oscar Lanford advanced a complete 
proof~\cite{L75}: he iterates the BBGKY hierarchy,  regarding extra collisions with a given cluster as perturbation. 
The resulting series is estimated uniformly for times $t$ with $|t| < T$, where $T$ is of the order of (1/5)-th of the mean free time. For such a short time span one can thus rely on the term-by-term convergence of the series. In the Boltzmann-Grad limit the series corresponds to the iteration of the Boltzmann equation regarding the collision operator as perturbation. In summary the Boltzmann equation is obtained as a limit when starting from Newton's equations of motion. The restriction to short times is an unavoidable feature of the perturbative expansion. To extend the validity of the result to longer times remains as an, apparently very difficult, open problem. 

But, even accepting the restriction to a finite kinetic time, Lanford's theorem leaves many issues without answer. 
Can smooth potentials be included? There are no explicit error bounds. Viewed probabilistically the Boltzmann
equation is a law of large numbers. In fact, consider the number density on the one-particle phase space for a 
single many-particle trajectory. This approximates the number density computed from the Boltzmann equation, with 
a probability very close to 1 relative to the initial non-equilibrium measure. Is there then a central limit 
type result relative to this law of large numbers?

The book under review presents a proof of Lanford's theorem with precise bounds allowing both 
an explicit rate of convergence for hard spheres, and an extension to the case of smooth potentials. 
In the latter direction a first step was achieved already in the 1975 Berkeley 
Ph.D. thesis by F. King \cite{Ki75}. As first observed by H. Grad \cite{G58}, one has to write down a version of the BBGKY 
hierarchy which uses completed collisions. The differential form used conventionally in physics text 
books is of no use when inserted into the perturbation series. The properly modified BBGKY hierarchy contains couplings to all higher order correlations. But King proved the same uniform bound as Lanford and also that the terms beyond next order 
vanish in the limit $\e \to 0$. It was then tacitly assumed that the term-by-term convergence could be accomplished by inspection. As the book demonstrates, this is far from the truth. Also, the explicit estimates require a much more detailed analysis
even in the hard sphere case. 

Besides providing useful background material, the bulk part of the book is concerned with solving an $n$-particle 
scattering problem for arbitrary but fixed $n$. To understand the nature of this problem, we briefly have to outline the 
structure of the perturbation series.

The starting point is a Hamiltonian  system consisting of $N$ particles which interact through a pair potential,
\be
\frac{dx_i}{dt}= v_i\;, \hspace{1cm} \frac{dv_i}{dt}= - \frac{1}{\e}\sum_{1 \leq j\neq i\leq N}\nabla\Phi\left(\frac{x_i-x_j}{\e}\right)\;,\nn
\ee
where $(x_i(t),v_i(t))\in \mathbb{R}^d\times\mathbb{R}^d$, 
$i=1,\ldots,N$,  denote the position and velocity of particle $i$ at time $t$, $\Phi$ 
is a given two-body potential with range $1$, and the physical dimension is $d=2,3$. 
Particles are indistinguishable, they move on straight lines with 
constant velocity and are deflected at a mutual distance $\e$. The singular case of hard spheres (formally, $\Phi(x)
= \infty$ if $|x| < 1$ and $\Phi(x)=0$ if $|x| > 1$) corresponds to the instantaneous elastic reflection.

We shall fix $N\e^{d-1} =~1$.
Writing $Z_N = (z_1,\ldots,z_N)$, $z_i = (x_i,v_i)$, the $N-$particle probability distribution is a function $f_N =~f_N(t,Z_N)$,
invariant under permutations of the particle labels, with evolution governed by the Liouville equation. In other words, 
$f_N(t)$ is obtained by transporting the prescribed initial distribution $f_N^0$ along the solutions of the 
above system of ODEs. Integrating $f_N$ over all the position and velocity variables but the first $s$, one defines the
$s$-particle marginal $f^{(s)}_N$. What we are interested in is the probability to find an arbitrary particle at position
$x$ and with velocity $v$ at time $t$, \textit{i.e.} $f^{(1)}_N(t,x,v)$. For $N$ large, its evolution is expected to be well described by
the Boltzmann equation
\be
\partial_t f + v \cdot \nabla_x f = \int_{\mathbb{S}_1^{d-1}} \int_{\mathbb{R}^d}\, [f'f_1' -ff_1] \, b(v-v_1,\o)\, dv_1 d\o,
\nn
\ee
where $\mathbb{S}_1^{d-1}$ is the unit sphere in $\mathbb{R}^d$,
$f = f(v)$, $f'=f(v')$, $f'_1 = f(v'_1)$ and $f_1 = f(v_1)$ with
$v' = v + \o\cdot(v_1-v)\o$, $v'_1 = v_1 - \o\cdot(v_1-v)\o$.
The collision kernel $b(w,\o)/|w|$ is the differential cross-section of the two-body scattering.

With the understanding that conditions on the potential $\Phi$ still 
need to be specified, a mathematical statement can be given as 
follows.\medskip\\
{\bf Theorem.} {\em Let $f_0: \mathbb{R}^{2d}\to \mathbb{R}^+$ be
a continuous probability density such that \\Ê$\|f_0\, \exp(\b |v|^2 / 2)\|_{L^\infty}<+\infty$ for some $\b>0$.
Let $E_\e(Z_s)$ be the $s$-particle energy,
\be
E_\e(Z_s) = \sum_{1\leq i \leq s} \frac{|v_i|^2}{2} + \sum_{1 \leq i<k\leq s} \Phi\left(\frac{x_i-x_k}{\e}\right)\;.\nn
\ee
Consider an initial ``asymptotically independent'' distribution of
particles such that, for some $\m>0$, the $s$-particle marginals satisfy
$|f^{(s),0}_N(Z_s)| \leq e^{\m s} e^{-\b E_\e(Z_s)}$ uniformly in $Z_s,s,N$,
and $f^{(s),0}_N \to f_0^{\otimes s}$ as $N\to \infty$, locally uniformly outside 
the diagonals of physical space, $\{x_i = x_k\}$. Then, there is a time $T >0$ such that for $t \in [0,T]$
\be
\lim_{\substack{N\rightarrow \infty\\ N\e^{d-1}=1}}f^{(1)}_N(t) = f(t)\;,\nn
\ee
where $f(t)$ is the solution of the Boltzmann equation with initial datum $f_0$.}\medskip

 A crucial byproduct
of the proof is that, in the same limit, $f^{(s)}_N(t) \to f(t)^{\otimes s}$ for any fixed $s$. In particular, the 
asymptotic factorization of the state propagates for short times.
The book provides a proof of the result, covering simultaneously the hard sphere potential
and a certain class of smooth, short range repulsive potentials. The convergence is weak in the sense that
one integrates against test functions with respect to the velocity variables, for any $Z_s$ outside 
diagonals of the physical space. In the case of hard spheres, a rate $\mathcal{O}(\e^\a)$ is shown to propagate in time, for any 
$\a < (d-1)/(d+1)$ and Lipschitz $f_0$.

The classical strategy resorts to the full set of evolution equations for the family of marginals 
$(f^{(s)}_N)_{1 \leq s \leq N}$: the BBGKY hierarchy. The rigorous derivation of these equations
and the short time control, uniform in $N$, of their solutions, are discussed by dealing separately with hard
spheres and smooth potentials (respectively parts II and III of the book). One faces indeed rather
orthogonal difficulties. In the hard sphere case, a nontrivial discussion is needed to conciliate
an integro-differential system of equations with the singular character of the dynamics. The $N$-particle flow
is defined only away from a set of Lebesgue measure zero, while the collision operators are defined by 
integrals on manifolds of codimension 1. In the proper weak formulation the hierarchy reads
\be
\pa_t f_N^{(s)} + \LL_s f_N^{(s)} = \CC _{s,s+1}f_{N}^{(s+1)}\;, \nn
\ee
where $ \LL_s = \sum_{1\leq i \leq s} v_i \cdot \nabla_{x_i}$ with boundary 
conditions identifying pre- and post-collisional velocities, and
\be
\CC _{s,s+1} f_{N}^{(s+1)}= (N-s)\e^{d-1} \sum_{i=1}^s \int_{\mathbb{S}_1^{d-1}\times\mathbb{R}^d}
\n \cdot (v_{s+1}-v_i) G_{<i,s+1>}
f_{N}^{(s+1)}(\cdot,x_i+\e\n,v_{s+1})\,d\n \ dv_{s+1}\;.\nn
\ee
Here $G_{<i,s+1>}$ is an eventually negligible constraint ensuring that particle $s+1$, with position $x_i+\e\n$,
keeps distance larger than $\e$ from all the other particles, except $i$. Repeated iteration of the 
time-integrated hierarchy produces Lanford's perturbative expansion. We refer to \cite{GSRT12E,Sp91,Si14} for the derivation of the 
perturbative setting. 

In case of a smooth interaction potential
the scattering is not instantaneous, but space-time delocalized and the equations change
drastically. The transport operator $\LL_s $ contains the additional term 
$-\sum_{1 \leq i\neq j \leq s} (1/\e)\nabla\Phi((x_i-x_j)/\e) \cdot \nabla_{v_i}$. The collision operator involves
also velocity derivatives and the particle $s+1$ runs over the interior of a sphere of radius $\e$ around $x_i$.
The iterated expansion becomes cumbersome, since binary collisions are mixed up with many-body
collisions. To overcome this problem, following King, one introduces truncated marginals,
\be
\tilde f_N^{(s)}(t,Z_s) = \int_{\mathbb{R}^{2d(N-s)}} f_N(t,Z_s,z_{s+1},\ldots,z_N) 
\prod_{\substack{ i\in\{1,\cdots,s\} \\ j\in \{s+1,\ldots,N\}} } \mathbbm{1}_{|x_i-x_j| > \e}
\, dz_{s+1} \ldots dz_N\;. \nn
\ee
These are asymptotically equivalent to the $f_N^{(s)}$. Moreover, in the hierarchy 
satisfied by $(\tilde f_N^{(s)})_{1 \leq s \leq N}$, the collision operator has the
same structure of the hard sphere $\CC_{s,s+1}$, plus corrections containing higher order marginals
and describing interactions of order three, four, \textit{etc.}, which a priori  can be shown to be negligible
in the Boltzmann-Grad limit (``cluster estimates'').
The resulting perturbation series has the form
\bea
\tilde f_N^{(s)}(t,Z_s) = \sum_{k=0}^\infty \, \int_0^t\int_0^{t_1}\cdots\int_0^{t_{k-1}}\,
\mathbf{H}_s(t-t_1)\,\CC_{s,s+1}\, \mathbf{H}_{s+1}(t_1-t_2)\, \CC_{s+1,s+2} \nn\\
\hfill \ldots\, \mathbf{H}_{s+k}(t_k)\,  \tilde f_{N}^{(s+k),0} \, dt_k \ldots dt_1 \,+ \mathcal{O}(\e)\,\;,\nn
\eea
where $\mathbf{H}_s$ is the $s$-particle flow associated to $\Phi$ and 
$\tilde f_{N}^{(s+k),0}$ are the initial data. The $\mathcal{O}(\e)$-error reduces to zero only when returning to the hard sphere case.

If $f(t)$ is a solution to the Boltzmann equation, then the family of tensor products 
$f^{(s)}(t,Z_s):= f(t,z_1)f(t,z_2) \ldots f(t,z_s)$ satisfies formally an analogous representation 
(Boltzmann hierarchical expansion), with the operators $\mathbf{H}_s$ and $\CC_{s,s+1}$ 
replaced respectively by the free flow operator and the hierarchical Boltzmann operator.
This discussion is included in parts II-III of the book, together with the above derivations and a priori
estimates, the notion of quasi-independence assumed for the time-zero state, and a precise formulation
of main results in terms of hierarchies, see Theorems 8 and 11.

Once the absolute convergence of the series expansion for $\tilde f_N^{(s)}$ is established for small times,
one is left with showing the term-by-term convergence to the Boltzmann series. This is the subject of part IV.
To understand the nature of the problem one has to look in detail at the structure of the generic 
term of the expansion.
The term $k=0$ is simple transport over the backward $s$-particle dynamics. The higher order terms 
are integrals over the trajectories of a special, fictitious, backward dynamics (in the book called ``pseudo-trajectories''), 
in which $k$ particles are adjoined to the first $s$ ones sequentially at times $t_1> t_2> \ldots> t_k$. 
Each new particle enters the evolution by appearing in a collision configuration, \textit{i.e.} at distance $\e$ from one of the already existing particles, its ``progenitor''. 
The precise way to add particles is specified by the 
integration variables appearing in the definition of the collision operator $\CC_{s,s+1}$. 
Between $t_{s+q}$ and $t_{s+q+1}$, the backward dynamics $\mathbf{H}_{s+q}$ is applied.
If the velocities of the new particle and its progenitor at time $t_{s+q}$ are post-collisional, then the
pair will scatter under the action of $\mathbf{H}_{s+q}$.
The initial datum $\tilde f_{N}^{(s+k),0}$ is integrated over the time-zero values of these special 
characteristics. As a consequence, the problem can be rephrased as a result of convergence 
in the proper space of trajectories. Namely, the dynamics associated to the (modified) BBGKY
perturbation series has to converge to the one of the Boltzmann series.
The main difference between the two is that in the Boltzmann flow the only  
scattering events retained are those which happen at the addition of new particles at  the times 
$t_1, t_2, \ldots, t_k$. 
In particular, each pair of particles collides at most once. Conversely, in the BBGKY flow other 
collisions, usually referred to as recollisions, may occur. Indeed, the heart of the proof consists in showing that the recollisions have a negligible measure in the space of pseudo-trajectories.

In the bulk of the book under review methods are developed which provide a quantitative estimate of the recollision set. One can see this 
as a scattering problem: the recollision described by an operator $\CC$ refers to the event that one of the two 
involved particles ends up hitting a third particle of the flow in the backward dynamics. This defines 
``untypical'' values of the integration variables in $\CC$, which can be geometrically characterized 
and estimated. Such an estimate is rendered delicate by the fact that it requires a certain stability. The 
collision $\CC$ itself introduces in fact additional deviations from the Boltzmann dynamics, due to a non-zero
collision time and the associated spatial displacement. Furthermore, one must be able
to iterate the process from $\CC_{s+q}$ to $\CC_{s+q+1}$. 
These difficulties are dealt with by a careful characterization of the
set of untypical trajectories and a clever use of the dispersivity
of the free flow.
The technique applies in the same way to hard spheres and to the considered class of smooth
potentials, where it leads to a rate of convergence depending implicitly on $\Phi$.

The results presented are a highly sophisticated completion of a program started some time ago.
In kinetic theory, the conventional lore is that the Boltzmann equation is applicable for ``any'' potential which decays faster than Coulomb. For potentials of infinite range, no matter how rapid a decay, the current methods fail because the total cross-section diverges. But even for a finite range
and smooth potential there will be initial conditions where the two-body collision time 
is infinite and the differential cross-section could be unbounded. The potentials covered in the book, see Sect. 15.2., have to satisfy a convexity assumption, which guarantees a bounded differential cross-section
away from grazing collisions. Subsequent studies on the subject followed~\cite{PSS13}, which use 
geometrical arguments to control singular cross-sections and long collision times, thereby enlarging the class of admissible potentials.
A more refined quantitative analysis for the hard sphere system
provides information on cumulants through estimates of many-recollision events~\cite{PS14}.
Recollision  trajectories have also to be controlled  over very long kinetic times for deriving  Brownian motion of a tracer particle immersed in a rarified hard sphere gas close to equilibrium \cite{BGSR13}, see also \cite{BNPP14} for diffusion in the low density Lorentz gas.

The presentation of the book is tuned so to mostly address an audience of mathematicians
working on partial differential equations.  Boltzmann's 1872 effort,  to conciliate microscopic time-reversible dynamics with increase of entropy and trend to equilibrium, still remains 
as a source of challenging mathematical problems.

\center{\textsc{References}}

\begingroup
\renewcommand{\section}[2]{}

\endgroup

\hfill \textsc{{\large Sergio Simonella}}

\smallskip

\hfill \textsc{WIAS Berlin}

\bigskip

\hfill \textsc{{\large Herbert Spohn}}
 
\smallskip
 
\hfill \textsc{TU M\"{u}nchen}

\bigskip

\hfill \texttt{simonell@wias-berlin.de}

\hfill \texttt{spohn@ma.tum.de}

\end{document}